# Ultra-low magnetic damping in $Co_2Mn$- based Heusler compounds: promising materials for spintronic

C. Guillemard[1,2], S. Petit-Watelot[1], L. Pasquier[1], D. Pierre[1], J. Ghanbaja[1], J-C. Rojas-Sànchez[1], A. Bataille[3], J. Rault[2], P. Le Fèvre[2], F. Bertran[2], S. Andrieu[1*]

[1] Institut Jean Lamour, UMR CNRS 7198, Université de Lorraine, 54506 Vandoeuvre lès Nancy, France

[2] Synchrotron SOLEIL-CNRS, L'Orme des Merisiers, Saint-Aubin, BP48, 91192 Gif-sur-Yvette, France

[3] Laboratoire Léon Brillouin, IRAMIS, CEA Saclay, 91191 Gif sur Yvette, France

*Email corresponding author: stephane.andrieu@univ-lorraine.fr

Abstract

The prediction of ultra-low magnetic damping in $Co_2MnZ$ Heusler half-metal thin-film magnets is explored in this study and the damping response is shown to be linked to the underlying electronic properties. By substituting the Z elements in high crystalline quality films ($Co_2MnZ$ with Z=Si, Ge, Sn, Al, Ga, Sb), electronic properties such as the minority spin band gap, Fermi energy position in the gap and spin polarization can be tuned and the consequence on magnetization dynamics analyzed. The experimental results allow us to directly explore the interplay of spin polarization, spin gap, Fermi energy position and the magnetic damping obtained in these films, together with *ab initio* calculation predictions. The ultra-low magnetic damping coefficients measured in the range $4.1\ 10^{-4} – 9\ 10^{-4}$ for $Co_2MnSi$, Ge, Sn, Sb are the lowest values obtained on a conductive layer and offers a clear experimental demonstration of theoretical predictions on Half-Metal Magnetic Heusler compounds and a pathway for future materials design.

## I - INTRODUCTION

As a result of many theoretical and experimental advances, spintronic, electronics that use both the charge and spin of the electron, is progressing. Predictions of many phenomena like high magnetoresistance MgO-based magnetic tunnel junction [1,2], magnetization reversal by spin-transfer torques (STT) [3,4], magnetization reversal by spin-orbit torques (SOT) [5] or all-optical switching (AOS) by direct laser excitation [6] offer possibilities to design magnetoresistive random access memories, magnetic sensors and novel logic devices. Even more the search for systems with high conversion efficiency of charge to spin or spin to charge current conversion [7] are being explored for fundamental understanding of spin-transport and for applications in low-energy-consumption devices. Most spintronic device consists of thin-film heterostructures where interesting physics emerges at the interfaces [8]. For continued progress, magnetic materials with specific and dedicated properties are needed, such as a high Curie temperature and an appropriate magnetic anisotropy for thermal stability [9], a high spin polarization at the Fermi energy to obtain fully spin-polarized currents, and a small magnetic damping to easily generate magnetization precession. All of these properties are desirable for STT, SOT and AOS based devices [9] but also in spin-wave-based devices, an emergent research field called magnonics [10].

However it is increasing challenging to achieve low magnetic damping in metallic magnetic materials. The magnetic damping reflects the ability of the magnetization to precess around an effective magnetic field. Dissipation occurs due to interactions with the environment, the precession amplitude decreases and the oscillating magnetization returns to align with the effective field. This damping process is characterized by the phenomenological Gilbert damping coefficient $\alpha$ within the Landau-Lifshitz-Gilbert (LLG) formalism [11-14]. For many emerging spintronic and magnonic application, this is particularly important in low-power applications that exploit magnetic dynamics such as STT switching where the switching current is directly proportional to $\alpha$ [15]. This is also true for SOT devices where a precessing magnetization generates a charge current in a metal and *vice et versa* [9,16].

While low damping parameters are often obtained in ferrimagnetic insulating oxides such as yttrium-iron garnet (YIG) where $\alpha=7.35 \times 10^{-5}$ can be observed in the bulk [17], magnetic metals typically have much higher damping where Fe-V alloys having a damping around $2 \times 10^{-3}$ [18, 19] was considered state of the art for a thin film. Recently there was progress in getting damping as low as $\alpha=2.1 \times 10^{-3}$ in Fe-Co alloys [20]. However, a broader class of materials where ultralow magnetic damping in metallic magnets emerges as half-metal magnetic (HMM) behavior. Such materials emerged in 1983 by De Groot *et al.* [21]

who reported predictions of peculiar electronic property of the NiMnSb half Heusler compound. Its electronic band structure was predicted to be in between a metal and an insulator. For the majority spin, responsible for the macroscopic magnetization, this material is a metal since electronic states are available around the Fermi energy. However, for minority spins, there is a gap around the Fermi energy. This peculiar property was called half-metallic magnetic (HMM) behavior. The NiMnSb material is thus a metal for majority spins whereas it is an insulator (at 0K) for minority spins. Such properties lead to a full spin-polarization at the Fermi energy, making this material an excellent candidate for spin current generation. Furthermore, additional theoretical studies performed on HMM materials highlighted another physical property of major importance in spintronic: their magnetic damping coefficients were predicted to be extremely low compared to other conductive materials (a factor 100 below in the $10^{-5}$ to $10^{-4}$ range) [22-24].

In HMM materials, extremely low magnetic damping coefficients are predicted due to the following reason. The electronic band structure imposes no density of states for minority spin. This spin channels exchange is thus forbidden and leads to continuous precession [22]. In practice, other dissipation processes are possible leading to non-zero damping coefficient, but even with taking them into account damping coefficient as low as $10^{-5}$ are predicted [22,24]. The precession damping is thus much smaller in a HMM than in a regular ferromagnetic material (with non-zero density of states at the Fermi energy for both spin channels). The HMM materials are thus very promising materials for applications.

After the paper of De Groot [21], HMM properties were theoretically predicted for many Heusler compounds [25-27]. However, it took a significant effort for experimentalists to obtain a direct verification. Indeed, the direct evidence of a spin gap in the minority spin channel was reported only very recently in $Co_2MnSi$ [28, 29]. On the other hand, as small magnetic damping coefficients were reported for several Heusler compounds, the measured values still remained in the $10^{-3}$ range [30-35] which is at least 10 times larger than prediction and comparable with epitaxial FeV alloys [19] which are not HMM. In 2016, we measured for the first time a damping coefficient in the $10^{-4}$ range [29] ($7.10^{-4}$ in $Co_2MnSi$, which was confirmed by another group in 2018 [36]). However, the fact that the magnetic damping values reported in the literature dedicated to HMM are often higher than $10^{-3}$ is puzzling. In fact, a recent theoretical study reports that the magnetic damping can strongly varies according to chemical disorder in the unit cell [24]. Thus, one big challenge when growing $X_2YZ$ full-Heusler thin films is to be as close as possible to the exact stoichiometry and achieving the chemically-ordered $L2_1$ phase as the outstanding properties of Co-based

Heusler compounds are most often predicted for this L2$_1$ phase. However, Heusler alloys can crystallize in several phases with a lower chemical ordering [37] without changing the atomic sites in the lattice. The most encountered disordered phase is the B2 one where Y and Z atoms are randomly arranged each other, leading to a primitive unit cell instead of the FCC cell ($Fm\overline{3}m \rightarrow Pm\overline{3}m$). Up to now, the extent to which the chemical disorder affects the physical properties is not clear. On one hand, *ab initio* calculations [24,38,39] and experiments [39,40] show that the physical properties (Curie temperature, cell parameter, magnetic moment, magnetic damping constant and spin polarization at E$_F$) are not drastically different between the L2$_1$ and B2 phases and the half-metallic spin gap should be conserved. On the other hand, the correlation between the degree of chemical ordering and the physical properties suffers a lack of experimental evidence since it is utterly complex to manipulate the degree of chemical ordering without adding other parameters to take into account (like introducing doping or reducing the crystallographic quality of the layer).

In this work, we present a systematic study of Co$_2$MnZ Heusler thin films epitaxially grown by molecular beam epitaxy (MBE) with Z=Al, Si, Ga, Ge, Sn, Sb. The chemical ordering inside the Heusler FCC lattice is examined by using *in situ* Reflection High Energy Electron Diffraction (RHEED) and *ex situ* Transmission Electron Microscopy (TEM). The spin polarization at the Fermi energy was determined by spin-resolved photoemission spectroscopy (SR-PES) experiments performed on the CASSIOPEE beamline at SOLEIL synchrotron source. Finally, the damping coefficients were measured by using FerroMagnetic Resonance (FMR) in a perpendicular geometry (dc magnetic field applied perpendicular to the film plane). Ultra-low damping (in the range 4. 10$^{-4}$ to 9 10$^{-4}$) is observed here for at least 4 of these 6 materials and is discussed according to the spin polarization determined experimentally and compared to theoretical predictions.

## II – EXPERIMENTAL RESULTS

*Samples growth and structural characterization* - All the films were grown by using Molecular Beam Epitaxy. The Heusler films stoichiometry was accurately controlled by using quartz microbalances (see appendix). *In situ* Electron diffraction (RHEED) performed all along the growth process allows us to control the epitaxial process and the chemical ordering in the Heusler unit cell. Indeed, if chemical ordering occurs, additional RHEED streaks should be observed along the Heusler [110] azimuth compared to the A2 or B2 phase [41]. In figure 1 are shown the typical RHEED patterns obtained after the growth process as well as the corresponding STEM-HAADF images. All the films shows additional streaks along [110]

RHEED patterns and STEM-HAADF images show the correct positioning of the Z element meaning that the films structure is compatible with the $L2_1$ structure, except $Co_2MnAl$ for which a B2 structure takes place, as reported in the literature [30].

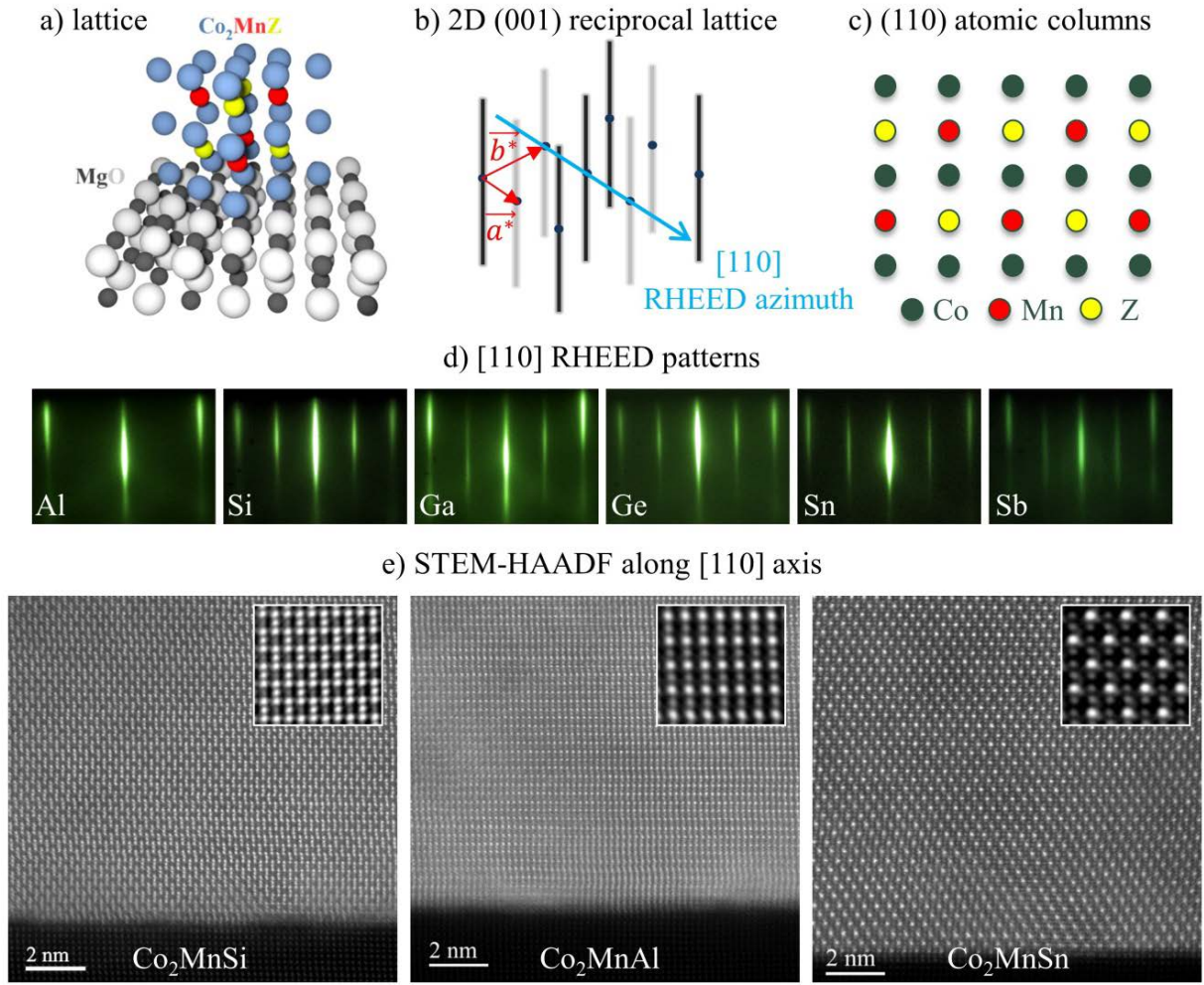

*Figure 1: a) Scheme of the epitaxial relationship between $Co_2MnZ$ films and MgO substrate, b) $Co_2MnZ$ (001) reciprocal lattice explored by electron diffraction (RHEED), c) Atomic columns along the [110] azimuth for the $L_{21}$ ordering, d) RHEED patterns along [110] for the $Co_2MnZ$ series, e) STEM-HAADF micrographs along [110] zone axis for the $Co_2MnSi$, $Co_2MnAl$ and $Co_2MnSn$ films (zoom in inset). Note the small weight of Si (14electrons) compared to Sn (50 electrons). The Z atoms are at the correct position in the cell as confirmed by RHEED and TEM except for $Co_2MnAl$ where a mixing of Mn and Si are observed leading to a B2 structure (no half streaks on the RHEED pattern and no contrast between the Mn and Z columns by microscopy).*

*Spin polarization* – Since SR-PES is a surface technique, the films were grown in a MBE chamber coupled to the synchrotron beamline (see appendix and ref. [29, 42]). The PES for each spin channel and the resulting Spin Polarization (SP) are shown in figure 2 for the Co$_2$MnZ series studied here. The results are discussed as a function of the Z element with the semiconductor terminology type III, IV and V and for simplicity we note the different Heusler compounds as Co$_2$MnZ$_{type}$.

For the Co$_2$MnZ$_{IV}$ compounds (Z=Si/Ge/Sn) the SR-PES spectra show a similar PES shape which is expected regarding the constant number of valence electrons in each compound. Importantly, one should note that they manifest the same transitions depicted in our previous paper on Co$_2$MnSi [29], where two features are explained. First, the loss of spin polarization at E$_F$ is caused by the presence of polarized surface states split by exchange coupling (noted S$_1$ for majority spin and S$_2$ for minority spin in figure 2). Indeed, these S$_1$ and S$_2$ transitions completely disappear when covering the surface by an atomic layer and the high spin polarization is thus observed up to the Fermi energy (we tested MgO, Mn and MnSi in [29]). It should be noted that these surface states are resonant with photon energy [29] so that there are observed only in the photon energy range 25-45 eV depending on the compounds (and are not observed in ref.28). We observed exactly the same behavior for Ge and Sn. However, there is a significant difference between Co$_2$MnSi when compared to Co$_2$MnGe and Co$_2$MnSn. The minority spin gap is large in Co$_2$MnSi, whereas it decreases in Co$_2$MnGe and Co$_2$MnSn. So the beginning of the spin gap is observed in Co$_2$MnSi (around -0.4 eV) despite the surface state contribution around E$_F$, but this is no longer the case for Ge and Sn. In other words, the PES experiments including the surface states do not allow one to obtain the bulk spin polarization. To access it, one can change the photon polarization from P to S. We previously showed that the spin gap can thus be observed up to E$_F$ in Co$_2$MnSi [29]. We consequently performed the same experiment on Co$_2$MnGe and Co$_2$MnSn. Results are shown in figure 2. Like in Co$_2$MnSi using a S photon polarization, the suppression of the transitions coming from the surface states allows us to see the spin gap in Co$_2$MnGe but more interestingly, a full spin polarization is now observed confirming Co$_2$MnGe is HMM. The (pseudo) spin gap is also observed for Co$_2$MnSn. Even though it is much larger than in figure 2 using P polarization this compound is not fully spin polarization but has a SP of 80%.

For Co$_2$MnZ$_{III}$ compounds, the number of valence electrons is lower than in type IV so E$_F$ should decrease. This is actually observed and the former surface states are above E$_F$ and not occupied so the SP is maximum at E$_F$. Therefore, the maximum of spin polarization visible on figure 2 for the Co$_2$MnZ$_{III}$ compounds nearly corresponds to the beginning of the minority spin gap. The SP was observed to be close to 60% for Co$_2$MnAl whereas a full SP was reached for Co$_2$MnGa.

For the $Co_2MnZ_V$ = Sb compound, a quite small SP is observed (around 35%) but one important feature has to be taking into account. Sb is well known to strongly segregate during MBE growth [43]. This segregation was also observed in Sb-based Heusler compounds like NiMnSb [44] or $Co_2TiSb$ [45]. We directly observed this segregation using Auger spectroscopy. Since PES is a surface technique, the electronic properties of such a Sb enriched surface are included with the underneath $Co_2MnSb$ PES contribution and may significantly affect the measured SP.

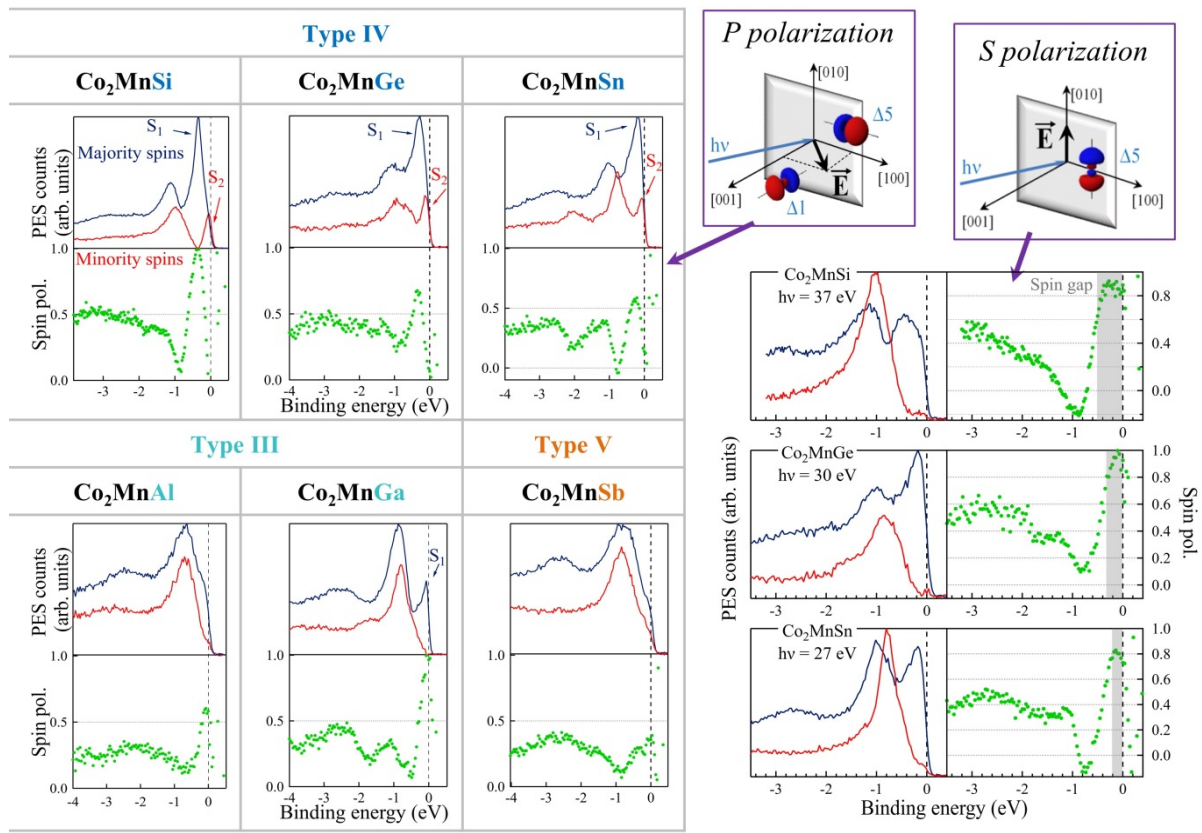

*Figure 2: Spin-Resolved photoemission spectra measured for the $Co_2MnZ$ series. An incoming photon energy of 37 eV is used for Z =Si, Al, Ga, Sb, 30 eV for Z=Ge and 27 eV for Z=Sn. We changed the incoming photon energy because of surface states, resonant in photon energy, which depends on the $Z_{IV}$ element in the compound. The left graphs were obtained using the P photon polarization. Surface transitions are observed and noted $S_1$ (majority spin channel) and $S_2$ (minority spin channel). $S_1$ and $S_2$ are responsible for the loss of spin polarization at $E_F$ (see text). Besides of that, spin gaps are observed for $Co_2MnGa$ and $Co_2MnSi$. The right graphs were obtained using the S polarization of the photons on $Co_2MnSi(001)$, $Co_2MnGe(001)$ and $Co_2MnSn(001)$. The $S_1$ and $S_2$ transitions coming from surface states are now forbidden leading to a clear observation of the minority spin gap.*

*Magnetic damping measurements* - The magnetic damping properties of these films were extracted by using FMR measurements performed on the samples grown on the CASSIOPPE beamline (after capping them with Au) but also on samples grown in our MBE at Institut Jean Lamour. FMR experiment allows studying the precession dynamic of the electronic magnetic moment [11-14,46,47] in ferromagnetic materials (see appendix for details). Typical VNA-FMR measurements performed on $Co_2MnGe$ (left column) and $Co_2MnSi$ (right column) are shown in figure 3. Figure 3-a corresponds to the $S_{11}$ reflection coefficient [48] for 1.9 T and 2.38 T of magnetic flux density, respectively. For each field value, the real and imaginary parts of the dynamic susceptibility are fitted simultaneously (*i.e.* using the same parameters to enhance the fitting reliability) to extract the position of the resonance frequency peak (figure 3-b) and its Half Width at Half Maximum (HWHM) (figure 3-c). At $f = 0$, the curve crosses the axis at $H = M_{eff} \approx M_S$ for small magnetic anisotropy, which is the case in these layers [49,50]. The measured effective magnetic moments per formula unit are reported in table I for the whole $Co_2MnZ$ series and are compared to theoretical values. Our experimental results tend to follow the $M_S = \Lambda - 24$ μB/f.u Slater-Pauling rules [51] where $\Lambda$ is the total number of valence electron in the Heusler structure. Finally, the linewidth slopes in figure 3-c give damping values equal to 5.3 $10^{-4}$ for $Co_2MnGe$ and 4.6 $10^{-4}$ for $Co_2MnSi$ at 290K (6.1 $10^{-4}$ and 4.1 $10^{-4}$ at 8K), the lowest of the $Co_2MnZ$ series (Figure 3-d). Moreover, inhomogeneous linewidth values $\Delta f_0$ in $Co_2MnGe$ and $Co_2MnSi$ (respectively 23.6 and 14.5 MHz) are very small, almost comparable with bulk magnetic insulators such as YIG. Such small values imply excellent homogeneity of the magnetic properties (hence a high crystal quality) in our films. Finally, one should note that this kind of FMR measurement leads to an effective value of the damping larger than the intrinsic value for the material [20,52,53]. Indeed, the effective damping measured here gathers extrinsic contributions such as the radiative damping, the eddy current damping and the spin-pumping contributions. Thus the magnetic damping values reported in table I include these extrinsic contributions and correspond to an upper limit of the true values. Recently, Shaw *et al.* [35] demonstrated a method to remove the extrinsic damping contributions in FMR measurements and get a better estimation of the true magnetic damping in $Co_2MnGe$ (measured damping 1.5 $10^{-3}$ which becomes smaller than $10^{-3}$ after corrections). Thus the theoretical values are likely found to be lower than experimental ones without correction as reported here. Even with these extrinsic contributions, the measured damping values of the whole $Co_2MnZ$ series are in the ultra-low damping range (four out of six in the $10^{-4}$ range and two below $2.10^{-3}$) in agreement with theoretical predictions.

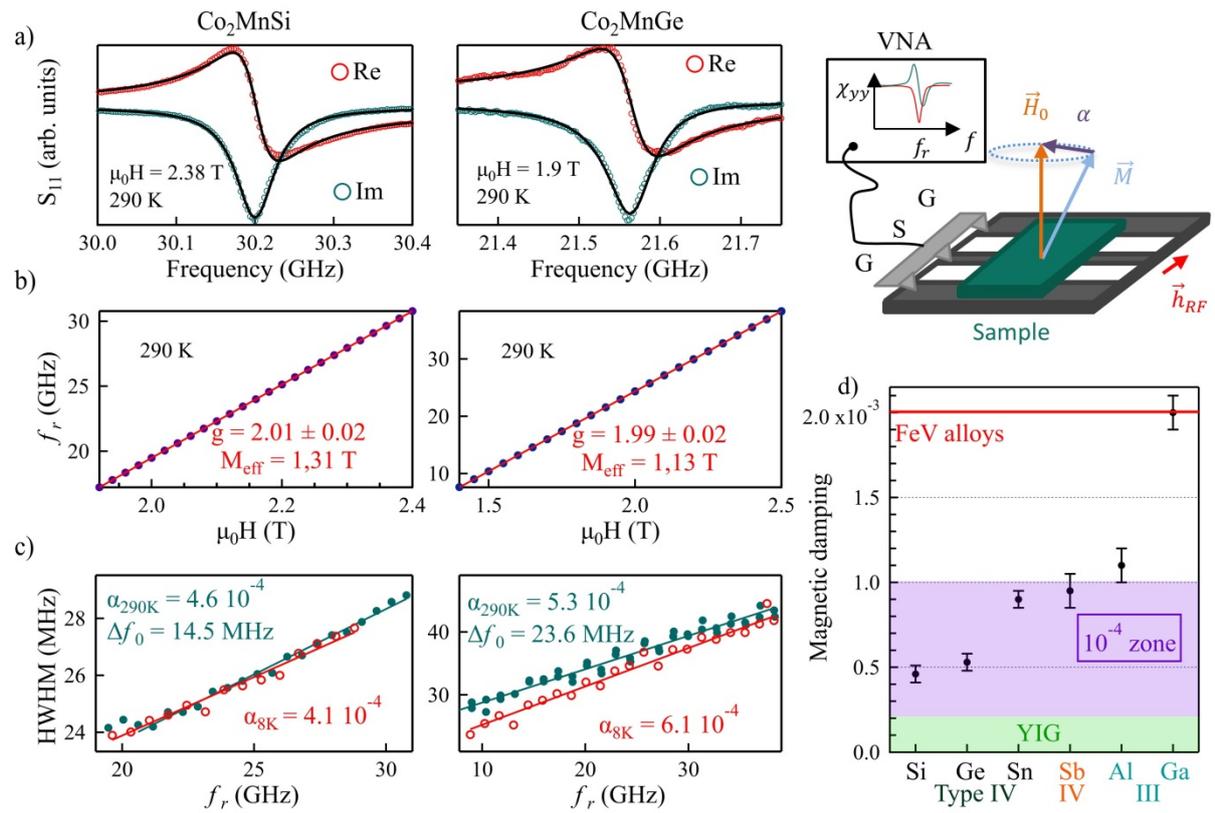

*Figure 3: Perpendicular FMR on $Co_2MnGe$ (left) and $Co_2MnSi$. Left -a) $S_{11}$ parameter of the scattering matrix, b) dependence of the FMR with field, c) evolution of the linewidth versus FMR. A review of the measured damping for the $Co_2MnZ$ series is also presented (right). The best magnetic damping obtained up to now in conductive FeV alloys thin films and bulk insulating YIG using our set-up are shown as a comparison.*

## III - DISCUSSION

All the experimental results extracted from this work are reported in table I together with theoretical predictions found in the literature. Looking first the PES results obtained on $Co_2MnZ_{IV}$ compounds, the observed full spin polarization for Si and Ge is consistent with theoretical calculations regarding the calculated width of the spin gap and the position of the Fermi energy in the middle of the gap. For $Co_2MnSi$ band structure calculations, the Fermi energy is located right in the middle of the spin gap [24, 54] and a large gap is obtained (0.8 eV [54], 0.41 eV [24,55]). Our results are consistent with these theoretical predictions since the measured (half-) gap between the minority spin valence band to $E_F$ is around 0.4 eV (figure 2). In the case of $Co_2MnGe$, the calculated gap is smaller (0.58 eV [56]) and $E_F$ is closer to the minority valence band, again consistent with our observations (half-gap around 0.25 eV – see figure 2). Finally, regarding $Co_2MnSn$ bands calculations, Kandpal *et al*. [56] found $E_F$ in the gap and close to the valence band (so HMM) whereas Ishida *et al*. [55] found it 0.06 eV below the top valence band (so not HMM). Our results do not allow us to conclude about this point (HMM or not) since our measurements are broadened by temperature (300K). However, both scenarios are consistent with the limited SP reported here. To summarize, the SR-PES results on these type IV Heusler compounds are well described by *ab initio* calculations. The presence of this minority spin gap at the Fermi energy should reduce the magnetic damping due to the removal of one conduction channel responsible for scattering processes involving spin flip (in other words, it should decrease the energy dissipation through spin relaxation in the system). Our magnetic damping measurements are in agreement with this mechanism. Indeed, the larger the spin gap the smaller the magnetic damping. These large spin gaps in $Co_2MnSi$ and $Co_2MnGe$ also explain why the magnetic damping is rather independent on the temperature for these two compounds (thermal fluctuation energy lower than the spin gap even at room temperature). But our FMR result allows us to go further in the case of $Co_2MnSn$, since its very low magnetic damping strongly suggests that $Co_2MnSn$ is a true HMM as predicted by [56]. To conclude, our results highlight a clear correlation between the theoretical gap width and Fermi energy position with the experimental SP and magnetic damping values.

| | | **Our experiments (290 K)** | | | **Theory (0 K)** | | | | |
|---|---|---|---|---|---|---|---|---|---|
| | Z | $a$ (Å) | $M_{eff}$ ($\mu_B$ / f.u) | $\alpha$ | Max of SP | $M_S$ ($\mu_B$/f.u) | $\alpha$ | SP | $\Delta_{Gap}$ (eV) | $E_F - E_{VB}$ (eV) |
| **Type III** | Al | 5.76 | 4.4 | $1.1\ 10^{-3}$ (290 K) $1.15\ 10^{-3}$ (30 K) | 60 % | 4,3 [38] | - | 0.68 [56] | 0.66 [56] | 0 [55] |
| | Ga | 5.77 | 5.4 | $2\ 10^{-3}$ (290 K) $1.8\ 10^{-3}$ (30 K) | 100 % | 4.09 [57] | - | 0.67 [56] | 0.3 [56] | 0 [55] |
| **Type IV** | Si | 5.65 | 5.1 | $4.6\ 10^{-4}$ (290 K) $4.1\ 10^{-4}$ (8 K) | 100 % | 4,94 [51] | $6\ 10^{-5}$ [22] | 100 [56] [24] | 0.41 [24] 0.81 [54] | ~ 0.2 [24] 0.33 [54] |
| | Ge | 5.75 | 5 | $5.3\ 10^{-4}$ (290 K) $6.1\ 10^{-4}$ (8 K) | 100 % | 4,94 [51] | $1.9\ 10^{-4}$ [21] | 100 [53] [55] | 0.58 [55] 0.54 [53] | 0.07 [60] 0.03 [53] |
| | Sn | 6.00 | 5.6 | $9.\ 10^{-4}$ (290 K) $10^{-3}$ (8 K) | 81 % | 4,98 [51] | $7\ 10^{-4}$ [59] | 0.77 [56] | 0.41 [56] 0.17 [55] | 0 [56] -0.06 [55] |
| **Type V** | Sb | 5.96 | 5.1 | $9.6\ 10^{-4}$ (290 K) | 36 % | 6 [58] | - | 100 [57] | 0.65 [58] 0.33 [55] | 0.56 [58] 0.39 [55] |

*Table I: review of the experimental results obtained by FMR and SR-PES and comparison with available theoretical results.*

Regarding bands calculations on $Co_2MnZ_{III}$ = Al or Ga considering the $L2_1$ phase, the theoretical SP at $E_F$ are respectively equal to 68 and 67% respectively [56]. Full SP is not reached because a small minority spin DOS remains at the Fermi energy. Strictly speaking, these compounds are not predicted to be true HMM. However, this former DOS is so small that a strong SP is still calculated. To account for this peculiar behavior, theoreticians use the term "pseudo gap" [57]. The SR-PES results shown in figure 2 for the B2 ordered $Co_2MnAl$ are in good agreement with this theoretical prediction since a SP of 60 % is obtained at $E_F$. It is also consistent with the theoretical work of B. Pradines *et al.* [24], explaining that the spin polarization at $E_F$ should stay unchanged between the $L2_1$ and B2 chemical phases. However, the situation is different for $Co_2MnGa$. We observed a full SP in clear disagreement with theory. Our experimental results show that $E_F$ is located in a true band gap, but very close to the minority spin valence band. As a consequence, this leads to a low magnetic damping (compared to regular ferromagnetic layers) but higher than type IV compounds due to the proximity of $E_F$ with the valence band.

For the $Co_2MnSb$ compound, theoretical calculations lead to a large spin gap (0.65 eV) with $E_F$ very close to the empty minority spin conduction band (0.09 eV below) in [58] or even in the conduction band in [55]. The experimental results are puzzling. As our PES results are not consistent with a HMM behavior, a very low magnetic damping is however observed consistent with a HMM behavior. In fact, the low SP can be explained by considering Sb segregation at the surface that does not affect the bulk SP. We conclude that $Co_2MnSb$ is likely HMM in the bulk, but its magnetic damping is higher than for type IV materials because of the $E_F$ proximity with the minority conduction band. These observations finally point out that our SR-PES analysis is pertinent to get bulk properties for (001) $Co_2MnSi$, Ge, Sn, Ga, and Al but not for (001) $Co_2MnSb$.

## IV - CONCLUSIONS

This work is a clear experimental demonstration that the unique electronic band structure of Half-Metal Magnetic materials leads to ultra-low magnetic damping. This study allows us to determine several key-points to achieve ultra-low damping. The most important ingredients are a large minority spin band gap, and a suitable position of the Fermi energy in the bandgap, that is not too close to minority spin valence and conduction bands. To achieve this behavior requires growing high-quality crystalline films with control of the stoichiometry. The measured magnetic damping values of the series match qualitatively with the spin polarization at $E_F$ predicted by the theoretical calculations and the SP measured by SR-PES without surface states. First, $Co_2MnSi$ and $Co_2MnGe$ have Fermi energy inside a large half-metallic gap and present ultra-low magnetic damping values below $6.10^{-4}$ (although the intrinsic damping will be even lower). Second, $Co_2MnSn$ and $Co_2MnSb$, with Fermi energy very close to a minority spin band (valence band for the former, conduction band for the latter) have magnetic damping values in between ($6.10^{-4} \leq \alpha \leq 10^{-3}$). And third, $Co_2MnZ_{III}$ (Al, Ga) are not predicted to be pure HMM and their damping values are obtained above the rest of the series with $10^{-3} \leq \alpha \leq 2.10^{-3}$. Nevertheless, the case of $Co_2MnGa$ is puzzling since we observed a full spin polarization whereas the magnetic damping coefficient is the highest in the series. Finally, the ultra-low magnetic damping we obtained in $Co_2MnSi$ ($4.1\ 10^{-4}$) and $Co_2MnGe$ ($5.3\ 10^{-4}$) have never been observed in a conductive material. One should note that in devices usually grown by sputtering the crystalline quality is not as good as in epitaxial films. However, the spin gap is predicted in the whole Brillouin Zone so there is no fundamental reason for not getting ultra-small damping in sputtered Heusler-based devices. Such ability to get easy precession of the magnetization offers extraordinary opportunities for getting more efficient spintronic devices, like Spin-Torque FMR, Spin Pumping FMR, Optical switching or magnonic based devices.

**ACKNOWLEDGMENT**

This work was supported partly by the french PIA project "Lorraine Université d'Excellence", reference ANR-15-IDEX-04-LUE. Some UHV experiments were performed using equipment from the TUBE—Davm funded by FEDER (EU), ANR, the Region Lorraine and Grand Nancy. We acknowledge Eric E. Fullerton from the Center for Memory and Recording Research (University of California San Diego-USA) for his critical reading of the manuscript.

**APPENDIX**

*Samples growth and structural characterization* - The substrates used to get single-crystalline Heusler films are MgO(001) due to the small misfit between MgO and $Co_2MnZ$ layers considered here. All the films were grown by using Molecular Beam Epitaxy. Co, Si and Ge were evaporated using electron guns, whereas Mn, Al, Ga, Sn and Sb by using Knudsen cells. Epitaxial films were obtained directly on MgO for Z=Al, Si, Ga, Ge. As this process was not successful with Z=Sb, $Co_2MnSb$(001) films were obtained on 10nm V(001) buffer layer grown on MgO. The Heusler films were all 20 nm thick. The starting temperature for the growth was fixed to 450°C (measured by a thermocouple beside the sample holder) and the Heusler films were thus heated up to 750°C after the growth to improve the chemical ordering in the lattice and the surface quality. Auger spectroscopy is systematically performed after the growth process and allowed us to verify that no surface contamination occurs (no O and C detected). The stoichiometry is a crucial point to get the best electronic and magnetic properties of these compounds. For this purpose, the fluxes of each element were calibrated by using a quartz microbalance located at the sample's location. The fluxes variations during the process are observed to be below 2%. To achieve the good stoichiometry, the Co flux is fixed to 2. $10^{14}$ at/cm$^2$/s and the Mn and Z elements fluxes to $10^{14}$ at/cm$^2$/s. Considering these fluxes and a sticking coefficient on the quartz equal to 1 for all these materials, the time to complete a layer is expected to be equal to 3.0 seconds (for $Co_2MnSi$). We measured exactly this completion time by using RHEED intensity oscillations performed during the growth. Moreover, the desired total thickness of the films (fixed by using the fluxes and the growth duration) is in perfect agreement with the final thickness determined by XRD reflectivity measurements. RHEED performed all along the growth process also allows us to verify that the Heusler lattice grows with the epitaxial relationship MgO [100] (001) // $Co_2MnZ$ [110] (001), meaning that the Heusler single-crystal is turned by 45° compared to the MgO lattice as shown in figure 1. As RHEED only gives information about the surface, θ-2θ XRD experiments were performed ex situ. All the films shows (111) peaks by XRD meaning that the films structure is compatible with the L2$_1$ structure, except $Co_2MnAl$ for which a no (111) peak was observed which is compatible with the B2 structure, as reported in the literature [30]. Finally, STEM-HAADF investigations were carried out using a JEM - ARM 200F Cold FEG TEM/STEM operating at 200 kV and equipped with a spherical aberration (Cs) probe and image correctors (point resolution 0.12 nm in TEM mode and 0.078 nm in STEM mode).

*Spin-resolved photo-emission spectroscopy (SR-PES)*: SR-PES experiments were performed on the CASSIOPEE beamline at SOLEIL synchrotron. The set-up and analysis process are detailed in [42] (beamline description) and [29] (study of Co$_2$MnSi). The spin-resolution is obtained using a Mott detector with an overall energy resolution of 150 meV in the photon energy range 30-40 eV used in this study. As discussed in ref. [29], our experimental conditions allow us to explore around 80% of the Brillouin Zone according to the photon energy range used in this study. As the films magnetization is always in-plane in our samples, the films were first magnetized *in situ* before the photoemission measurement by applying an in-plane magnetic field equal to 200 0e sufficient to saturate the magnetization (the coercive fields in our samples are always below 100 Oe). As the SR-PES measurement has to be performed without any magnetic field, the spin polarization extracted from raw PES spectra is obtained at remanence. The remanence is systematically measured on the same films *ex situ* after capping the samples with 5 nm thick gold. The remanence was observed to vary from 80 to 100% depending on the Co$_2$MnZ compound. The true spin polarizations shown in this paper are thus obtained by correcting the SR-PES spectra from remanence.

*Magnetic damping measurements*: Like in NMR, the precession occurs when an oscillating magnetic field of small amplitude is generated perpendicular to the magnetic moment equilibrium which corresponds, in a ferromagnet, to the magnetization direction. This equilibrium position is imposed by the effective field derived from magnetic free energy density that includes exchange, dipolar interactions and magnetocrystalline anisotropy energies. The magnetic damping is, by definition, inversely related to the lifetime of the precession. The damping value is thus extracted by looking at the linewidth of the resonance peak in frequency. The larger the linewidth is, the higher the damping and the shorter the precession motion. FMR experiments were performed in the *perpendicular geometry* where the static magnetic field is applied out of the plane of the film in order to avoid extrinsic broadening of the linewidth due to the 2-magnons scattering [46,47]. The RF magnetic field is generated, thanks to a Vector Network Analyser (VNA-FMR), in a coplanar waveguide (ground-signal-ground geometry) on top of which the sample is placed face down. Measurements are performed in reflection geometry. The physical parameter extracted from these experiments is the S$_{11}$ coefficient of the scattering matrix of the line, from which the dynamic susceptibility of the magnetic layer is extracted [47]. In the case of a perpendicular geometry, the Kittel law [11] (equation 1) becomes linear and so does the evolution of the peak's linewidth versus its own resonance frequency. The slope of this curve is directly the magnetic damping value (equation 2):

$$f_r = \gamma_0(H - M_S + H_{k\perp}) = \gamma_0(H - M_{eff}) \quad (1)$$

$$\Delta f = 2\alpha\gamma_0(H - M_{eff}) + 2\Delta f_0 = 2\alpha f_r + 2\Delta f_0 \qquad (2)$$

where $f_r$ refers to the resonance frequency, $\gamma_0$ the gyromagnetic ratio of the electron, $H$ the magnetic field strength, $M_S$ the magnetization, $H_{k\perp}$ the perpendicular magnetic anisotropy, the effective magnetization $M_{eff} = M_S - H_{k\perp}$, $\Delta f$ the full width at half maximum, $\Delta f_0$ the inhomogeneous half linewidth and $\alpha$ the Gilbert magnetic damping. The shift in frequency of the resonance peak versus field is imposed by the gyromagnetic ratio $\gamma_0$ of the electron, which should be equal to 28 GHz/T in the case of a of a pure delocalized ferromagnetic model (free electron model). However, in a solid, this ratio can be different mainly due to spin-orbit coupling since $\gamma_0$ is proportional to the Landé $g$-factor. The fitted slopes in figure 3-c give $\gamma_0 = 27.9$ and 28.2 GHz/T corresponding to a Landé $g$-factor equal to 1.99 and 2.01 (very close from the non-interacting electron case, as expected for 3d compounds).

References

[1]  W. H. Butler, X.-G. Zhang, T.C. Schulthess, and J.M. MacLaren, Spin-dependent tunneling conductance of Fe|MgO|Fe sandwiches, Phys. Rev. B **63**, 054416 (2001).

[2]  J. Mathon, and A. Umerski, Theory of tunneling magnetoresistance of an epitaxial Fe/MgO/Fe(001) junction, Phys. Rev. B **63**, 220403(R) (2001).

[3]  J. C. Slonczewski, Conductance and exchange coupling of two ferromagnets separated by a tunneling barrier, J. Magn. Magn. Mater. **159**, L1 (1996).

[4]  L. Berger, Emission of spin waves by a magnetic multilayer traversed by a current, Phys. Rev. B **54**, 9353 (1996).

[5]  I.M Miron, K. Gatello, G. Gaudin, P-J. Zermatten, M.V. Costache, S. Auffret, S. Bandiera, B. Rodmacq, A. Schuhl, and P. Gambardella, Perpendicular switching of a single ferromagnetic layer induced by in-plane current injection, Nature, **476**, 189 (2011).

[6]  C-H. Lambert, S. Mangin, B. S. D. Ch. S. Varaprasad Y. K. Takahashi, M. Hehn, M. Cinchetti, G. Malinowski, K. Hono, Y. Fainman, M. Aeschlimann, and E. E. Fullerton, All-optical control of ferromagnetic thin films and nanostructures, Science **345**, 1337 (2014).

[7]  J-C. Rojas-Sánchez, N. Reyren, P. Laczkowski, W. Savero, J-P. Attané, C. Deranlot, M. Jamet, J-M. George, L. Vila, and H. Jaffrès, Spin pumping and inverse spin Hall effect in platinum: the essential role of spin-memory loss at metallic interfaces, Phys. Rev. Lett. **112**, 106602 (2014).

[8]  F. Hellman, A. Hoffman, Y. Tserkovnyak, G.S.D. Beach, E.E.Fullerton, C. Leighton, A.H. MacDonald, D.C. Ralph, D.A. Arena et al, Interface-induced phenomena in magnetism Rev. Mod. Phys. **89**, 025006 (2017).

[9]  I.L. Prejbeanu, S Bandiera, J Alvarez-Hérault, R C Sousa, B Dieny, and J-P Nozières, Thermally assisted MRAMs: ultimate scalability and logic functionalities, J. Phys. D: Appl. Phys. **46**, 074002 (2013).

[10]  A. V. Chumak, and H. Schultheiss, Magnonics: spin waves connecting charges, spins and photons, J. Phys. D: Appl. Phys. **50**, 300201 (2017).

[11]  C. Kittel, On the Theory of Ferromagnetic Resonance Absorption, Phys. Rev. **73**, 155 (1948).

[12]  H. Suhl, Theory of the magnetic damping constant, IEEE Transactions on magnetics, **34**, 1834 (1998).

[13]  V. Kamberský, Spin-orbital Gilbert damping in common magnetic metals, Phys. Rev. B **76**, 134416 (2007).

[14]  K. Gilmore, Y. U. Idzerda, and M. D. Stiles, Identification of the Dominant Precession-Damping Mechanism in Fe, Co, and Ni by First-Principles Calculations, Phys. Rev. Lett. **99**, 027204 (2007).

[15]  D.C. Ralph, and M.D. Stiles, Spin transfer torques, J. Mag. Mag. Mat., **320**, 1190 (2008).

[16]  K-S. Lee, S-W. Lee, B-C. Min, and K-J. Lee, Threshold current for switching of a perpendicular magnetic layer induced by spin Hall effect, Appl. Phys. Lett. **102**, 112410 (2013).

[17]   C. Hauser, T. Richter, N. Homonnay, C. Eisenschmidt, M. Qaid, H. Deniz, D. Hesse, M. Sawicki, S. G. Ebbinghaus, and G. Schmidt, Yttrium Iron garnet Thin Films with Very Low Damping Obtained by Recrystallization of Amorphous Material, Sci. Rep. **6,** 20827 (2016).

[18]   C. Scheck, L. Cheng, I. Barsukov, Z. Frait, and W. E. Bailey, Low Relaxation Rate in Epitaxial Vanadium-Doped Ultrathin Iron Films, Phys. Rev. Lett. **98**, 117601 (2007).

[19]   T. Devolder, M. Manfrini, T. Hauet, and S. Andrieu, Compositional dependence of the magnetic properties of epitaxial FeV/MgO thin films, Appl. Phys. Lett. **103**, 242410 (2013).

[20]   M. A.W. Schoen, D. Thonig, M. L. Schneider, T. J. Silva, H. T. Nembach, O. Eriksson, O. Karis, and J. M. Shaw, Ultra-low magnetic damping of a metallic ferromagnet, Nature Physics **12**, 839-842 (2016).

[21]   R. A. deGroot, F. M. Mueller, P.G. vanEngen, and K. H. J. Buschow, New class of materials: half-metallic ferromagnets, Phys. Rev. Lett. **50**, 2024 (1983).

[22]   C. Liu, C. K. A. Mewes, M. Chshiev, T. Mewes, and W.H. Butler, Origin of low Gilbert damping in half metals, Appl. Phys. Lett. **95**, 022509 (2009).

[23]   B. Pigeau, G. de Loubens, O. Klein, A. Riegler, F. Lochner, G.Schmidt, L.W.Molenkamp, V. S. Tiberkevich, and A. N. Slavin, Frequency-controlled magnetic vortex memory, Appl. Phys. Lett., **96**, 132506 (2010).

[24]   B. Pradines, R. Arras, I. Abdallah, N. Biziere, and L. Calmels, First-principles calculation of the effects of partial alloy disorder on the static and dynamic magnetic properties of $Co_2MnSi$, Phys. Rev. B **95**, 094425 (2017).

[25]   I. Galanakis, and P. Mavropoulos, Spin-polarization and electronic properties of half-metallic Heusler alloys calculated form first prinicples, J. Phys.: Condens. Matter **19**, 315213 (2007).

[26]   I. Kastnelson, V. Yu. Irkhin, L. Chioncel, A.I. Lichtenstein, and R.A. De Groot, Half-metallic ferromagnets: from band structure to many-body effects, Rev. Mod. Phys. **80**, 315 (2008).

[27]   T. Graf, C. Felser, and S. S. P. Parkin, Simple rules for the understanding of Heusler compounds, Prog. Solid State Ch. **39**, 1 (2011).

[28]   M. Jourdan, J. Minar, J. Braun, A. Kronenberg, S. Chadov, B. Balke, A. Gloskovskii, M. Kolbe, H. J. Elmers, G. Schonhense, H. Ebert, C. Felser, and M. Klaui, Direct observation of half-metallicity in the Heusler compound $Co_2MnSi$, Nat. Commun. **5**, 4974 (2014).

[29]   S. Andrieu, A. Neggache, T. Hauet, T. Devolder, A. Hallal, M. Chshiev, A.M. Bataille, P. Le Fèvre, and F. Bertran, Direct evidence for minority spin gap in the $Co_2MnSi$ Heusler compound, Phys. Rev. B **93**, 094417 (2016).

[30]   S. Trudel, O. Gaier, J. Hamrle, and B. Hillebrands, Magnetic anisotropy, exchange and damping in cobalt-based full-Heusler compounds: An experimental review, J. Phys. D: Appl. Phys. **43**, 19 (2010).

[31]   M Oogane, T Kubota, H Naganuma, and Y Ando, Magnetic damping constant in Co-based full heusler alloy epitaxial films, J. Phys. D: Appl. Phys. **48**, 164012 (2015).

[32] S. Husain, S. Akansel, A. Kumar, P. Svedlindh, and S. Chaudhary, Growth of $Co_2FeAl$ Heusler alloy thin films on Si(100) having very small Gilbert damping by Ion beam sputtering, Sci. Rep. **6**, 28692 (2016).

[33] I. Abdallah, B. Pradines, N. Ratel-Ramond, G. Benassayag, R. Arras, L. Calmels, J-F.Bobo, E. Snoeck, and N. Biziere, Evolution of magnetic properties and damping coefficient of $Co_2MnSi$ Heusler alloy with Mn/Si and Co/Mn atomic disorder, J. Phys. D: Appl. Phys. **50**, 035003 (2017).

[34] L Bainsla, R Yilgin, M Tsujikawa, K Z Suzuki, M Shirai, and S Mizukami, Low magnetic damping for equiatomic CoFeMnSi Heusler alloy, J. Phys. D: App. Phys. **51**, 495001 (2018).

[35] J. M. Shaw, E. K. Delczeg-Czirjak, E. R. J. Edwards, Y. Kvashnin, D. Thonig, M. A. W. Schoen, M. Pufall, M. L. Schneider, T. J. Silva, O. Karis, K. P. Rice, O. Eriksson, and H. T. Nembach, Magnetic damping in sputter-deposited $Co_2MnGe$ Heusler compounds with A2, B2, and $L2_1$ orders: Experiment and theory, Phys. Rev. B **97**, 094420 (2018).

[36] M. Oogane, A. P. McFadden, K. Fukuda, M. Tsunoda, Y. Ando, and C. J. Palmstrøm, Low magnetic damping and large negative anisotropic magnetoresistance in halfmetallic $Co_{2-x}Mn_{1+x}Si$ Heusler alloy films grown by molecular beam epitaxy, Appl. Phys. Lett. **112**, 262407 (2018).

[37] G. E. Bacon, and J. S. Plant, Chemical ordering in Heusler alloys with the general formula $A_2BC$ or ABC, J. Phys. F: Metal. Phys. **1**, 524-532 (1971).

[38] X. Zhu, E. Jiang, Y. Dai, and C. Luo, Stability, magnetic, & electronic properties of $L2_1$ and B2 phases in $Co_2MnAl$ Heusler alloy, J. Alloys Compd **632**, 528–532 (2015).

[39] A. Kumar, F. Pan, S. Husain, S. Akansel, R. Brucas, L. Bergqvist, S. Chaudhary, and P. Svedlindh, Temperature-dependent Gilbert damping of $Co_2FeAl$ thin films with different degree of atomic order, Phys. Rev. B **96**, 224425 (2017).

[40] A. Rajanikanth, D. Kande, Y. K. Takahashi, and K. Hono, High spin polarization in a two phase quaternary Heusler alloy $Co_2MnAl_{1-x}Sn_x$, J. Appl. Phys. **101**, 09J508 (2007).

[41] A. Neggache, T. Hauet, F. Bertran, P. Le Fèvre, S. Petit-Watelot, T. Devolder, Ph. Ohresser, P. Boulet, C. Mewes, S. Maat, J.R. Childress, and S. Andrieu, Testing epitaxial $Co_{1.5}Fe_{1.5}Ge$ electrodes in MgO-based MTJs, Appl. Phys. Lett. **104**, 252412 (2014).

[42] S. Andrieu, L. Calmels, T. Hauet, F. Bonell, P. Le Fèvre, and F. Bertran, Spectroscopic & transport studies of $Co_xFe_{1-x}$/MgO based MTJs, Phys. Rev. B, **90**, 214406 (2014).

[43] S. Andrieu, Sb adsorption on Si(111) analyzed by elippsometry and Reflexion High Energy Electron Diffraction – Consequences for Sb doping in molecualr beam epitaxy, J. Appl. Phys. **69**, 1366 (1991).

[44] P. Turban, S. Andrieu, B. Kierren, E. Snoeck, C. Teodorescu, and A. Traverse, Growth and characterization of single crystalline NiMnSb thin films and epitaxial NiMnSb / MgO / NiMnSb(001) trilayers, Phys. Rev. B **65**, 134417 (2002).

[45] J. K. Kawasaki, A. Sharan, L. I. M. Johansson, M. Hjort, R. Timm, B. Thiagarajan, B. D. Schultz, A. Mikkelsen, A. Janotti, and C. J. Palmstrøm, A simple electron counting model for half-Heusler surfaces, Sci. Adv. **4**, eaar5832 (2018).

[46] K. Lenz, H. Wende, W. Kuch, K. Baberscke, K. Nagy, and A. Jánossy, Two-magnon scattering and viscous Gilbert damping in ultrathin ferromagnets, Phys. Rev. B **73**, 144424 (2006).

[47] K. Zakeri, J. Lindner, I. Barsukov, R. Meckenstock, M. Farle, U. von Hörsten, H. Wende, W. Keune, J. Rocker et al, Spin dynamics in ferromagnets: Gilbert damping and two-magnon scattering, Phys. Rev. B **76**, 104416 (2007).

[48] C. Bilzer, T. Devolder, Joo-Von Kim, G. Counil, C. Chappert, S. Cardoso, and P. P. Freitas, Study of the dynamic magnetic properties of soft CoFeB films, J. Appl. Phys. **100**, 053903 (2006).

[49] Yu. V. Goryunov, N. N. Garif'yanov, G. G. Khaliullin, I. A. Garifullin, L. R. Tagirov, F. Schreiber, Th. Mühge, and H. Zabel, Magnetic anisotropies of sputtered Fe films on MgO substrates, Phys. Rev. B **52**, 13450 (1995).

[50] A. Butera, J. L. Weston, and J. A. Barnard, Ferromagnetic resonance of epitaxial $Fe_{81}Ga_{19}$(110) thin films, J. Mag. Mag. Mat. **284**, 17 (2004).

[51] I. Galanakis, P. H. Dederichs, and N. Papanikolaou, Slater-Pauling behavior and origin of the half-metallicity of the full-Heusler alloys, Phys. Rev. B **66**, 174429 (2002).

[52] M. A. W. Schoen, J. M. Shaw, H. T. Nembach, M. Weiler, and T. J. Silva, Radiative damping in waveguide-based ferromagnetic resonance measured via analysis of perpendicular standing spin waves in sputtered permalloy film, Phys. Rev. B **92,** 184417 (2015).

[53] Y. Li, and W. E. Bailey, Wave-Number-Dependent Gilbert Damping in Metallic Ferromagnets Phys. Rev. Lett. **116**, 117602 (2016).

[54] S. Picozzi, A. Continenza, and A. J. Freeman, $Co_2MnX$ (X = Si, Ge, Sn) Heusler compounds: An ab initio study of their structural, electronic, and magnetic properties at zero and elevated pressure, Phys. Rev. B **66**, 094421 (2002).

[55] S. Ishida, S. Fujii, S. Kashiwagi, and S. Asano, Search for Half-Metallic Heusler Compounds in $Co_2MnZ$ (Z = IIIb, IVb, Vb Element), J. Phys. Soc. Jpn. **64**, 2152-2157 (1995).

[56] H.C. Kandpal, G. H. Fecher, and C. Felser, Calculated electronic and magnetic properties of the half-metallic, transition metal based Heusler compounds, J. Phys. D: Appl. Phys. **40,** 1507–1523 (2007).

[57] K. Özdoğan, E. Şaşıoğlu, B. Aktaş, and I. Galanakis, Doping and disorder in the $Co_2MnAl$ and $Co_2MnGa$ half-metallic Heusler alloys, Phys. Rev. B **74**, 172412 (2006).

[58] H. M. Huang, S. J. Luo, and K. L. Yao, First-principles investigation of the electronic structure and magnetism of Heusler alloys CoMnSb and $Co_2MnSb$, Physica B **406,** 1368–1373 (2011).

[59] A. Pradines, R. Arras, and L. Calmels, Effects of partial B2, D03 and A2 disorders on the magnetic properties of the Heusler alloys $Co_2FeAl$, $Co_2MnSn$ and $Co_2MnAl$ for spintronic applications, in preparation.

[60] S. Ouardi, Electronic Structure and Physical Properties of Heusler Compounds for Thermoelectric and Spintronic Applications, Ph.D Thesis, Johannes Gutenberg-University at Mainz, Germany (2012).